\input harvmac

\input psfig
\newcount\figno
\figno=0
\def\fig#1#2#3{
\par\begingroup\parindent=0pt\leftskip=1cm\rightskip=1cm\parindent=0pt
\global\advance\figno by 1
\midinsert
\epsfxsize=#3
\centerline{\epsfbox{#2}}
\vskip 12pt
{\bf Fig. \the\figno:} #1\par
\endinsert\endgroup\par
}
\def\figlabel#1{\xdef#1{\the\figno}}
\def\encadremath#1{\vbox{\hrule\hbox{\vrule\kern8pt\vbox{\kern8pt
\hbox{$\displaystyle #1$}\kern8pt}
\kern8pt\vrule}\hrule}}

\overfullrule=0pt



\def\unlockat{\catcode`\@=11}
\def\lockat{\catcode`\@=12}

\unlockat

\def\newsec#1{\global\advance\secno by1\message{(\the\secno. #1)}
\global\subsecno=0\global\subsubsecno=0\eqnres@t\noindent
{\bf\the\secno. #1}
\writetoca{{\secsym} {#1}}\par\nobreak\medskip\nobreak}
\global\newcount\subsecno \global\subsecno=0
\def\subsec#1{\global\advance\subsecno
by1\message{(\secsym\the\subsecno. #1)}
\ifnum\lastpenalty>9000\else\bigbreak\fi\global\subsubsecno=0
\noindent{\it\secsym\the\subsecno. #1}
\writetoca{\string\quad {\secsym\the\subsecno.} {#1}}
\par\nobreak\medskip\nobreak}
\global\newcount\subsubsecno \global\subsubsecno=0
\def\subsubsec#1{\global\advance\subsubsecno by1
\message{(\secsym\the\subsecno.\the\subsubsecno. #1)}
\ifnum\lastpenalty>9000\else\bigbreak\fi
\noindent\quad{\secsym\the\subsecno.\the\subsubsecno.}{#1}
\writetoca{\string\qquad{\secsym\the\subsecno.\the\subsubsecno.}{#1}}
\par\nobreak\medskip\nobreak}

\def\subsubseclab#1{\DefWarn#1\xdef
#1{\noexpand\hyperref{}{subsubsection}%
{\secsym\the\subsecno.\the\subsubsecno}%
{\secsym\the\subsecno.\the\subsubsecno}}%
\writedef{#1\leftbracket#1}\wrlabeL{#1=#1}}
\lockat

\def\IL{\relax{\rm I\kern-.18em L}}
\def\IH{\relax{\rm I\kern-.18em H}}
\def\IR{\relax{\rm I\kern-.18em R}}
\def\IC{\relax\hbox{$\inbar\kern-.3em{\rm C}$}}
\def\IT{\relax\hbox{$\inbar\kern-.3em{\rm T}$}}
\def\IZ{\relax\ifmmode\mathchoice
{\hbox{\cmss Z\kern-.4em Z}}{\hbox{\cmss Z\kern-.4em Z}}
{\lower.9pt\hbox{\cmsss Z\kern-.4em Z}}
{\lower1.2pt\hbox{\cmsss Z\kern-.4em Z}}\else{\cmss Z\kern-.4em
Z}\fi}


\def\CT {{\cal T}}

\font\manual=manfnt \def\dbend{\lower3.5pt\hbox{\manual\char127}}

\def\IZ{\relax\ifmmode\mathchoice
{\hbox{\cmss Z\kern-.4em Z}}{\hbox{\cmss Z\kern-.4em Z}}
{\lower.9pt\hbox{\cmsss Z\kern-.4em Z}}
{\lower1.2pt\hbox{\cmsss Z\kern-.4em Z}}\else{\cmss Z\kern-.4em
Z}\fi}
\def\half {{1\over 2}}

\def\p{\partial}


\def\IZ{\relax\ifmmode\mathchoice
{\hbox{\cmss Z\kern-.4em Z}}{\hbox{\cmss Z\kern-.4em Z}}
{\lower.9pt\hbox{\cmsss Z\kern-.4em Z}}
{\lower1.2pt\hbox{\cmsss Z\kern-.4em Z}}\else{\cmss Z\kern-.4em
Z}\fi}
\def\IB{\relax{\rm I\kern-.18em B}}
\def\IC{{\relax\hbox{$\inbar\kern-.3em{\rm C}$}}}
\def\ID{\relax{\rm I\kern-.18em D}}
\def\IE{\relax{\rm I\kern-.18em E}}
\def\IF{\relax{\rm I\kern-.18em F}}
\def\IG{\relax\hbox{$\inbar\kern-.3em{\rm G}$}}
\def\IGa{\relax\hbox{${\rm I}\kern-.18em\Gamma$}}
\def\IH{\relax{\rm I\kern-.18em H}}
\def\II{\relax{\rm I\kern-.18em I}}
\def\IK{\relax{\rm I\kern-.18em K}}
\def\IP{\relax{\rm I\kern-.18em P}}
\def\IQ{\relax\hbox{$\inbar\kern-.3em{\rm Q}$}}

\def\inbar{\,\vrule height1.5ex width.4pt depth0pt}

\def\p{\partial}

\font\cmss=cmss10 \font\cmsss=cmss10 at 7pt
\def\IR{\relax{\rm I\kern-.18em R}}


\def\boxit#1{\vbox{\hrule\hbox{\vrule\kern8pt
\vbox{\hbox{\kern8pt}\hbox{\vbox{#1}}\hbox{\kern8pt}}
\kern8pt\vrule}\hrule}}
\def\mathboxit#1{\vbox{\hrule\hbox{\vrule\kern8pt\vbox{\kern8pt
\hbox{$\displaystyle #1$}\kern8pt}\kern8pt\vrule}\hrule}}


\def\inbar{\,\vrule height1.5ex width.4pt depth0pt}

\def\p{\partial}

\font\cmss=cmss10 \font\cmsss=cmss10 at 7pt
\def\IR{\relax{\rm I\kern-.18em R}}



\lref\mw{G. Moore and E. Witten, ``Integration over the $u$-plane in 
Donaldson theory'', Adv. Theor. Math. Phys. {\bf 1} (1997) 298, 
hep-th/9709193.}

\lref\sw{N. Seiberg and E. Witten, ``Electric-magnetic duality, monopole 
condensation, and confinement in ${\cal N}=2$ supersymmetric Yang-Mills
Theory'', Nucl. Phys. {\bf B 426} (1994) 19, hep-th/9407087. ``Monopoles, 
Duality and Chiral Symmetry Breaking in ${\cal N}=2$ supersymmetric QCD'',
Nucl. Phys. {\bf B 431} (1994) 484, hep-th/9408099.}

\lref\lns{A. Losev, N. Nekrasov, and S. Shatashvili, ``Issues in
topological gauge theory'', Nucl. Phys. {\bf B 534} (1998) 549, 
hep-th/9711108. ``Testing Seiberg--Witten solution'', in L. Baulieu et.
al. (eds.), {\it Strings, branes and dualities}, Kluwer Academic 
Publishers, Cargese, pp. 359, hep-th/9801061.}

\lref\mmone{M. Mari\~no and G. Moore, ``Integrating over the Coulomb branch
in ${\cal N}=2$ gauge theory'', Nucl. Phys. B (Proc. Suppl.) {\bf 68} (1998)
336, hep-th/9712062.}

\lref\mmtwo{M. Mari\~no and G. Moore, ``The Donaldson-Witten function for 
gauge groups of rank larger than one'', Commun. Math. Phys. {\bf 199} (1998)
25, hep-th/9802185.}

\lref\emm{J.D. Edelstein, M. Mari\~no and J. Mas, ``Whitham hierarchies, 
instanton corrections, and soft supersymmetry breaking in ${\cal N}=2$ 
$SU(N)$ super Yang-Mills theory'', Nucl. Phys. {\bf B 541} (1999) 671, 
hep-th/9805172.}

\lref\bu{J.D. Edelstein, M. G\'omez--Reino and M. Mari\~no, ``Blowup 
formulae in Donaldson--Witten theory and integrable hierarchies'',
Adv. Theor. Math. Phys. {\bf 4} (2000) in press, hep-th/0006113.}

\lref\marijapon{M. Mari\~no, ``The uses of Whitham hierarchies'', Progr. 
Theor. Phys. Suppl. {\bf 135} (1999) 29, hep-th/9905053.}

\lref\victoruno{V.M. Buchstaber, V.Z. Enolskii and D.V. Leykin, ``Kleinian 
functions, hyperelliptic Jacobians, and applications'', in {\it Reviews 
in Mathematics and Mathematical Physics} {\bf 10:2}, eds. S.P. Novikov and 
I.M. Krichever, London 1997.} 

\lref\nato{J.D. Edelstein and M. G\'omez-Reino, ``Integrable hierarchies 
in Donaldson-Witten and Seiberg-Witten theories'', hep-th/0010061.} 

\lref\takasaki{K. Takasaki, ``Integrable hierarchies and contact terms in 
$u$-plane integrals of topologically twisted supersymmetric gauge
theories'', Int. J. Mod. Phys. {\bf A 14} (1999) 1001, hep-th/9803217. 
``Whitham deformations and tau functions in ${\cal N} = 2$ supersymmetric 
gauge theories'', Prog. Theor. Phys. Suppl. {\bf 135} (1999) 53,
hep-th/9905224.}

\lref\bolza{O. Bolza, ``On the first and second logarithmic derivatives 
of hyperelliptic $\sigma$-functions'', Amer. J. Math. {\bf 17} (1895) 11.
``Proof of Brioschi's recursion formula for the expansion of the even 
$\sigma$--function of two variables'', Amer. J. Math. {\bf 21} (1899) 1.
``The partial differential equations for the hyperelliptic $\Theta$ and 
$\sigma$--functions'', Amer. J. Math. {\bf 21} (1899) 107. ``Remarks
concerning the expansions of the hyperelliptic $\sigma$--functions'',
Amer. J. Math. {\bf 22} (1900) 101.}

\lref\witteni{E. Witten, ``On $S$-duality in abelian gauge theory'',
Selecta Mathematica {\bf 1} (1995) 383, hep-th/9505186.}

\lref\sunmat{A. Hanany and Y. Oz, ``On the quantum moduli space of vacua 
of ${\cal N}=2$ supersymmetric $SU(N_c)$ gauge theories'', Nucl. Phys. {\bf 
B 452} (1995) 283, hep-th/9505075. P.C. Argyres, M.R. Plesser and A. 
Shapere, ``The Coulomb phase of ${\cal N}=2$ supersymmetric QCD'', Phys. 
Rev. Lett. {\bf 75} (1995) 1699, hep-th/9505100.}

\lref\egrmm{J.D. Edelstein, M. G\'omez--Reino, M. Mari\~no and J. Mas, 
``${\cal N}=2$ supersymmetric gauge theories with massive hypermultiplets 
and the Whitham hierarchy'', Nucl. Phys. {\bf B 574} (2000) 587,
hep-th/9911115.}

\lref\belo{E.D. Belokolos, A.I. Bobenko, V.Z. Enolskii, A.R. Its, and V.B. 
Matveev, {\it Algebro--geometric approach to nonlinear integrable
equations}, Springer--Verlag, 1994.}

\lref\kra{H.M. Farkas and I. Kra, {\it Riemann Surfaces}, Springer--Verlag,
1991.}

\lref\mmp{M. Mari\~no, G. Moore and G. Peradze, ``Superconformal invariance 
and the geography of four-manifolds'', Commun. Math. Phys. {\bf 205} (1999)
691, hep-th/9812055.}

\lref\klt{A. Klemm, W. Lerche and S. Theisen, ``Nonperturbative effective 
actions of ${\cal N}=2$ supersymmetric gauge theories'', Int. J. Mod. Phys.
{\bf A 10} (1996) 1029, hep-th/9505150.}

\lref\egrm{J.D. Edelstein, M. G\'omez--Reino and J.~Mas, ``Instanton 
corrections in ${\cal N}=2$ supersymmetric theories with classical gauge 
groups and fundamental matter hypermultiplets'', Nucl. Phys. {\bf B 561}
(1999) 273, hep-th/9904087.}

\lref\itep{A. Gorsky, A. Marshakov, A. Mironov and A. Morozov, ``RG 
equations from the Whitham hierarchy'', Nucl. Phys. {\bf B 527} (1998) 690, 
hep-th/9802007.}

\lref\edemas{J.D. Edelstein and J. Mas, ``Strong coupling expansion and 
the Seiberg--Witten--Whitham equations'', Phys. Lett. {\bf B 452} (1999) 69, 
hep-th/9901006.}

\Title{\vbox{\baselineskip12pt
\hbox{HUTP-00/A046}
\hbox{US-FT/20-00}
\hbox{RUNHETC-2000-49}
\hbox{hep-th/0011227}
}}
{\vbox{\centerline{Remarks on Twisted Theories with Matter}}}
\centerline{Jos\'e D. Edelstein$^{a}$, Marta G\'omez--Reino$^{b}$ and 
Marcos Mari\~no$^{c}$}

\bigskip
\medskip
{\vbox{\centerline{$^{a}$ \sl Lyman Laboratory of Physics, Harvard University}
\centerline{\sl Cambridge, MA 02138, USA}
\centerline{\it edels@lorentz.harvard.edu}}}

\bigskip
\medskip
{\vbox{\centerline{$^{b}$ \sl Departamento de F\'\i sica de Part\'\i culas,
Universidade de Santiago de Compostela} 
\centerline{\sl E-15706 Santiago de Compostela, Spain}
\centerline{\it marta@fpaxp1.usc.es}}}

\bigskip
\medskip
{\vbox{\centerline{$^{c}$ \sl New High Energy Theory Center, Rutgers 
University} 
\centerline{\sl Piscataway, NJ 08855, USA}
\centerline{\it marcosm@physics.rutgers.edu}}}

\bigskip
\bigskip
\noindent
We investigate some aspects of ${\cal N}=2$ twisted theories with matter 
hypermultiplets in the fundamental representation of the gauge group. A 
consistent formulation of these theories on a general four-manifold 
requires turning on a particular magnetic flux, which we write down 
explicitly in the case of $SU(2\ell)$. We obtain the blowup formula and 
show that the blowup function is given by a hyperelliptic $\sigma$--function
with singular characteristic. We compute the contact terms and find, as a 
corollary, interesting identities between hyperelliptic Theta functions.

\Date{21 November 2000} 

\listtoc \writetoc

\newsec{Introduction}

In the last few years, some work has been devoted to the study of an 
interesting interplay between supersymmetric gauge theories and their 
twisted counterparts, and integrable hierarchies (for a review, see 
\marijapon\nato). This interplay appears in a very natural way in the 
context of the so-called $u$-plane integral introduced by Moore and 
Witten \mw, and has been explored in some detail in 
\lns\mmtwo\itep\takasaki\emm. 
In a recent paper \bu, we have extended and clarified the role of integrable 
systems in twisted ${\cal N}=2$ theories by using the theory of hyperelliptic 
Kleinian functions. This allowed us to fully characterize the blowup function
of Donaldson--Witten theory with gauge group $SU(N)$ and, furthermore, to 
identify it as a $\tau$ function of a finite-gap solution (a multisoliton 
solution in the case of four-manifolds of simple type) of the KdV 
hierarchy. As a corollary of this result, we obtained a new expression 
for the contact terms ${\cal T}_{k,l}$ that appear in the low-energy 
twisted theory.

In this note we shall extend some of these results to the case of
twisted theories with matter hypermultiplets. This case has been 
comparatively less studied except in the case of gauge group $SU(2)$, 
which was considered in \mw\lns\mmone\ and has been shown to lead to new
results in four-manifold topology \mmp. An interesting feature emerging in
this framework, as shown in \mw, is that $SU(2)$ twisted theories with matter 
are only consistent on a general four-manifold if one turns on a magnetic 
flux. We shall give a generalization of this mechanism for gauge group 
$SU(2\ell)$ and matter in the fundamental representation. As in the 
$SU(2)$ case \mw, this magnetic flux is related to the second 
Stiefel--Whitney class of the four-manifold.

We shall present some new results for the blowup formulae which are
valid in any twisted theory with gauge group $SU(2\ell)$ (with or without 
matter). In particular, we identify the blowup function as a hyperelliptic
{\it fundamental} $\sigma$--function (whose characteristic is that of the
vector of Riemann constants). The characteristic being singular, we show 
that the leading contribution to the blowup function is of order 
$\ell^2$ in the ``times'' $t_i$, and we describe a procedure to expand it in
terms of the vacuum expectation values of local observables of the twisted 
theory up to arbitrary order.

We finally derive novel expressions for the contact terms corresponding to
descendant operators whose supporting two-cycles intersect. As a corollary 
of this analysis, we obtain a family of identities among
hyperelliptic $\Theta$--functions. The interpretation of these contact
terms within the framework of the Whitham hierarchy leads to a new
equation for the Seiberg--Witten effective prepotential.

\newsec{Twisted theories with matter}

In this section we shall consider the extension of the $u$-plane integral 
\mw\mmtwo\lns\ in Donaldson--Witten theory with gauge group $SU(N)$ when
matter in the fundamental representation is included. We show the
appearance of topological obstructions to define a monodromy invariant
$u$-plane integral, and the way in which they can be overcome when the rank
of the gauge group is odd. We conclude this section by presenting the
blowup formula.

\subsec{The $u$-plane integral}

The $u$-plane integral gives the answer for the generating functional of 
twisted ${\cal N}=2$ $SU(N)$ theories on four-manifolds $X$ with $b_2^+=1$.
The basic observables of the twisted theory are the Casimir operators of 
the gauge group,
\eqn\casimirs{
{\cal O}_k = {1\over k} {\rm Tr}\phi^k + {\rm lower \, order \, terms} ~,
~~~ k=2, \dots, N ~,}
whose vacuum expectation values $u_k = \langle {\cal O}_k \rangle$ are gauge
invariant coordinates for the Coulomb branch of the theory. Starting from 
the Casimirs, one can construct further (topological descendant) observables
associated to two-cycles $S\in H_2(X)$ on the four-manifold,
\eqn\descent{
I_k(S) = {1 \over k}\int_S {\rm Tr}(\phi^{k-1} F) 
+ \dots ~,}
the dots standing for superpartner contributions. The generating functional 
of correlation functions in the twisted theory is then defined as:
\eqn\genfun{
Z(p_k,f_k, S) = \Big\langle \exp \bigl[ \sum_{k=2}^N ( p_k {\cal O}_k + 
\sum_{i=1}^{b_2(X)}f_{k,i} I_k(S_i) )\bigr] \Big\rangle ,}
where the $S_i$, $i=1, \cdots, b_2(X)$, define a basis of $H_2(X)$. 

The generating functional \genfun\ can be explicitly evaluated by means of
the exact low-energy effective action of the ${\cal N}=2$ theory \sw. This 
action is encoded in a hyperelliptic curve $y^2=f(x)$, the so-called 
Seiberg--Witten curve. As explained in \mw\mmtwo, for generic values of the
hypermultiplet bare masses --such that there is no Higgs branch--, \genfun\ 
has two contributions: one comes from the Coulomb branch, $Z_{\rm Coulomb}$,
and the other comes from the vanishing locus ${\cal D}$ of the discriminant 
$\Delta$ of the Seiberg--Witten curve. $Z_{\rm Coulomb}$ is given by an 
integral over the Coulomb branch, ${\cal M}_{\rm Coulomb} = \IC^{N-1} -
{\cal D}$, the previously alluded $u$-plane integral. Its explicit 
expression turns out to be: 
\eqn\coulint{
Z_{\rm Coulomb} = \int_{{\cal M}_{\rm Coulomb}} [da  d\bar   a]~
A(u_k)^\chi B(u_k)^\sigma  e^{\sum p_k u_k  + \sum f_{k,i} f_{l,j} 
{\cal T}_{k,l}(S_i,S_j)} ~\Psi ~.}
The integrand of \coulint\ has various ingredients. First of all, there is
a gravitational measure first studied in \witteni\ in the pure $SU(2)$
case, generalized in \mmtwo\lns\ to simply-laced groups, and further
extended in \mw\ to the case of $SU(2)$ with matter hypermultiplets. In the
present case, the measure is given by:
\eqn\measure{
A(u_k)^\chi = \alpha^{\chi} \biggl( {\rm det} {\partial u_k \over \partial 
a^i} \biggr)^{\chi/2} ~, \,\,\,\,\,\ B(u_k)^\sigma = \beta^{\sigma} 
\Delta^{\sigma/8} ~,}
where $\chi$ and $\sigma$ are the Euler characteristic and the signature of
$X$, while $\alpha$ and $\beta$ are functions of the bare masses $m_f$ and
the dynamically generated scale $\Lambda$, that remain constant along
${\cal M}_{\rm Coulomb}$, and whose determination might be achieved by 
comparison with the Donaldson--Witten theory at short distances. The first 
factor in \measure\ involves the determinant of the matrix of periods, 
whereas the second one contains the discriminant of the hyperelliptic 
curve. The factor $\Psi$ is given by a complicated sum over a lattice 
$\Gamma$ that involves a generalized Siegel--Narain theta function. A 
vector $\vec\lambda$ of this lattice is of the form
\eqn\lams{
\vec \lambda = \vec \lambda_{\IZ} + \vec \xi ~,}
where
\eqn\integers{
{\vec \lambda}_{\IZ} = \sum_{i=1}^{N-1} \lambda_{\IZ}^i \vec \alpha_i ~,}
$\lambda_{\IZ}^i$ are elements of $H^2(X, \IZ)$, $\vec \alpha_i$ are the 
simple roots of $SU(N)$, and $\vec \xi$ is the magnetic flux that 
corresponds to the second Stiefel--Whitney class of the gauge bundle. 
For $SU(N)$, it has the form \mmtwo:
\eqn\fluxsi{
\vec \xi  = \sum_{i=1}^{N-1} p^i \vec w_i ~,}
where $p^i$ are fixed elements of $H^2(X, \IZ)$ that represent a choice of
the second Stiefel--Whitney class of the bundle $w_2(E)$, and $\vec w_i$ 
are the fundamental weights of $SU(N)$. The function $\Psi$ is 
the same appearing in the pure gauge case \mw\mmtwo. The dependence 
on the effective coupling in this function has the form
\eqn\sumlat{ 
\exp \biggl[ - i \pi {\overline \tau} _{ij} ({
\lambda}^i_+, { \lambda}^i_+) - i \pi \tau_{ij} ({ \lambda}^i_-, 
{ \lambda}^j_-) \biggr] ~.}
In \sumlat,
$\lambda_+ =(\lambda,\omega)$ is the self-dual part of the two-form $\lambda$,
constructed out of the unique anti-self-dual form $\omega \in H^{2,+}(X, 
{\IR})$ such that $\omega^2=1$, by means of the usual $(~,~)$ product in 
cohomology. Finally, $Z_{\rm Coulomb}$ contains contact terms ${\cal 
T}_{k,l}$ among the different two-observables. These contact terms have 
been studied in great detail in \mw\lns\mmtwo\itep\takasaki\emm\bu, and we 
will come back to them later on.

\subsec{Monodromy invariance}

As emphasized in \mw\mmtwo, in order to have a well-defined $u$-plane 
integral, the integrand of \coulint\ should be {\it invariant} under the 
monodromies associated to the hyperelliptic curve. In the $SU(2)$ theory 
with matter \mw, this requirement leads to a nontrivial condition for the 
magnetic flux. Let us presently analyze the same problem for twisted 
theories with gauge group $SU(N)$ and $N_f$ massive hypermultiplets in the 
fundamental representation. The hyperelliptic curve corresponding to the
infrared dynamics of the untwisted theory for $N_f \leq N$ is \sunmat\
\eqn\curve{
y^2 = P(\lambda,u_k)^2 - 4\Lambda^{2N-N_f} F(\lambda,m_f) ~,}
where $P$ is the characteristic polynomial of $SU(N)$
\eqn\chpol{
P = x^N - \sum_{k=2}^{N} u_{k} x^{N-k} ~,}
and $F$ endowes the dependence on the bare masses,
\eqn\funf{
F = \prod_{f=1}^{N_f} (x+m_f) ~.} 
The case $N < N_f < 2N$ can be similarly considered. We will show that,
provided $N$ is even, there is a particular value of the magnetic flux for
which the $u$-plane integral is invariant under the semiclassical monodromies.
Usually this is enough to guarantee the invariance under the strong 
coupling monodromies as well, so we will assume that this is the right 
choice to have a consistent $u$-plane integral.

The semiclassical monodromies that are specific of theories with matter,
are those associated to elementary quarks becoming massless. They have the 
following form \klt:
\eqn\monomatter{
a_{D,i} \longrightarrow a_{D,i} + (\vec\mu_I \otimes \vec\mu_I)_{ij} ~a^j ~,}
where $\vec \mu_I$,  $I=1, \cdots, N$, are the weights of the fundamental 
representation of $SU(N)$. This monodromy corresponds to encircling the 
semiclassical singularity $\vec a \cdot \vec \mu_I + m_f=0$. 
Under this monodromy, the effective couplings change as:
\eqn\coumat{
\tau_{ij} \longrightarrow \tau_{ij} + (\vec \mu_I)_i (\vec \mu_I)_j ~.}
The monodromy introduces a phase in the $u$-plane integral, which can be 
computed as in \mmtwo. The measure \measure\ contains a factor 
$\Delta^{\sigma/8}$. Going around the elementary quark singularity one 
picks a phase $\exp (\pi i \sigma/4)$. The other phase comes from the 
contribution \sumlat\ to the Siegel--Narain theta function, as a 
consequence of the shift in the effective gauge couplings under the 
monodromy,
\eqn\otraph{
\exp \Bigl[ -i\pi (\vec\lambda \cdot \vec\mu_I,\vec\lambda \cdot 
\vec\mu_I)\Bigr] ~.}
This has to cancel against the overall phase coming from the measure. In 
particular, the phase \otraph\ should be independent of $\vec\lambda$. It 
is easy to check that one has indeed the desired cancellation if $N$ is 
{\it even} and the magnetic flux is given by
\eqn\fluxchoice{
\vec\xi  = w_2(X) \vec\rho ~,}
where $\vec\rho$ is the Weyl vector ({\it i.e.} the sum of the fundamental 
weights), and $w_2(X)$ is the second Stiefel--Whitney class of the 
four-manifold $X$.
Notice that for $SU(2)$ one obtains, in the root basis, that $\xi_1 = 
w_2(X)/2$, which is precisely the magnetic flux found in \mw\ for theories 
with massive matter. To verify that \fluxchoice\ guarantees the 
cancellation, one has to use Wu's formula (which states that 
$(w_2(X),\alpha) \equiv \alpha^2$ mod $2$ for any two-cohomology class), 
the fact that $w_2^2(X) \equiv \sigma$ mod $8$, and that 
\eqn\wyl{
\vec\mu_I \cdot \vec\rho = {1 \over 2} (N-2I+1) ~.}
We will assume from now on that $N$ is even, $N = 2\ell$. It may be 
possible that there is a choice of $\vec\xi$ which makes the theory 
well-defined also for $N$ odd, but we have not found any. Notice that the 
need to choose a flux reflects the fact that in the twisted theory there are 
fields which are sections of the bundle $S^+ \otimes E$, where $S^+$ is 
the spinor bundle on $X$ and $E$ is the gauge bundle. In general, this 
product bundle does not exist, and this is what requires the choice of a 
nontrivial Stiefel--Whitney class for the gauge bundle. It would be 
interesting to understand \fluxchoice\ from this point of view, and this 
might give a hint of how to make the choice of $\vec\xi$ for $N$ odd (or 
prove that there is none).

\subsec{Blowup formula}

Using this information we can already compute the blowup formula as in 
\mw\mmtwo\lns. Consider the four-manifold ${\widehat X}$, obtained from $X$ 
by blowing up a point $p$, $\widehat X = {\rm Bl}_{p}(X)$. This means that 
there is a map $\pi:\widehat X \to X$ that is the identity everywhere except 
at $B = \pi^{-1}(p)$, where $B \in H_2({\widehat X})$ such that $B^2 = -1$. 
$B$ is called the class of the exceptional divisor. The anti-self-dual
two-form contribution to the $u$-plane integral in $\widehat X$ gets
modified to ${\widehat \lambda}_-^i = \lambda_-^i + n^i B$ with $n^i \in \IZ$.
Up to now, we have considered the blowup formula only when there is no 
magnetic flux through the exceptional divisor. As noticed in \mw\ for the 
$SU(2)$ case, the choice \fluxchoice\ actually forces us to shift the flux 
by $B \vec\rho$. This is due to the fact that $w_2(\widehat X) = w_2(X) +B$ 
(mod $2$). Thus, vectors $\vec{\widehat \lambda}$ of the lattice $\widehat
\Gamma$ corresponding to the $u$-plane integral of the blownup
four-manifold have the form
\eqn\buvec{
\vec{\widehat \lambda} = \vec\lambda + (\vec\alpha + \vec\rho) ~B = 
\vec\lambda + \sum_{i=1}^{N-1} (n^i + \sum_{j=1}^{N-1} (C^{-1})_j^{~i})
\vec\alpha_i ~B ~,}
$C$ being the Cartan matrix. This amounts to the appearance of a particular
characteristic $\sum_{j=1}^{N-1} (C^{-1})_j^{~i}$ in the factor \sumlat,
which is easily seen to be integer (half-integer) for even (odd) $i$. Thus,
the $\vec\beta$ characteristic of the blowup function becomes
\eqn\betach{
\vec\beta = \vec\Sigma = (1/2,0,1/2,\cdots,0,1/2) ~.}
The $\vec\alpha$ characteristic is the same as in the pure gauge case
\eqn\alphach{
\vec\alpha = \vec\Delta = (1/2,1/2,\cdots,1/2,1/2) ~,}
so that, finally, the blowup function of the massive twisted theory with 
gauge group $SU(2\ell)$ has a half-integer characteristic, which is even 
(odd) for even (odd) values of $\ell$. Aside of this important aspect, the
derivation of the blowup formula within the $u$-plane integral follows the 
same lines developed in \mw\mmtwo. In particular, there is a contribution
from the measure due to the fact that both the Euler characteristic and the
signature of the manifold are shifted by the blowup. The final outcome is 
that the generating functional \genfun\ for the blownup four-manifold can 
be written as
\eqn\bugenfun{\eqalign{
\widehat Z(p_k,f_k,t_k,S) = & \biggl\langle \exp\bigl[ \sum_{k=2}^{N}
(p_k {\cal O}_k + t_k I_k(B) + \sum_{i=1}^{b_2(X)} f_{k,i} I_k(S_i)) \bigr] 
\biggr\rangle_{\widehat{X},\vec\Sigma} \cr = &
\biggl\langle \exp\bigl[ \sum_{k=2}^{N} (p_k {\cal O}_k + 
\sum_{i=1}^{b_2(X)} f_{k,i} I_k(S_i))\bigr] ~\tau_{\vec\Sigma}(t_k|{\cal 
O}_k) \biggr\rangle_X ~,}}
where the blowup function $\tau_{\vec\Sigma}(t_i|u_k)$ is given by
\eqn\blowupf{
\tau_{\vec\Sigma}(t_i|u_k) = \biggl( {\rm det} {\partial u_k \over \partial 
a^j} \biggr)^{1/2}~ \Delta^{-1/8} ~\exp\biggl\{-\sum_{k,l=2}^N t_k t_l 
{\CT}_{k,l}\biggr\} ~\Theta[\vec\Delta,\vec\Sigma](\vec z| \tau) ~,} 
up to an overall factor which only depends on $m_f$ and $\Lambda$. The 
argument of the $\Theta$--function in \blowupf\ is 
\eqn\vector{
z_i = \sum_{k=2}^N {t_k \over 2\pi}{\partial u_{k} \over \partial a^i} ~.} 
As in the case of pure ${\cal N}=2$ Yang--Mills, the blowup can be 
interpreted as a local defect, and we 
expect the blowup function to be given by an expansion of the form
\eqn\expansion{
\tau_{\vec\Sigma} (t_i|{u}_k) = \sum_{n\ge 0}\sum_{i_1, \cdots, i_n} 
t_{i_1} \cdots t_{i_n} 
{\cal B}^{(n)}_{i_1 \cdots i_{n}}
(u_k,m_f,\Lambda) ~,}
where ${\cal B}^{(n)}_{i_1 \cdots i_{\ell}}
(u_k,m_f,\Lambda)$ are homogeneous polynomials, the degree of their variables
being given by their mass dimensions. 

\newsec{The blowup function}

In this section we shall study in detail many aspects of the blowup function
corresponding to twisted theories with matter. We first identify the fixed
characteristic of the $\Theta$--function as a singular one. This fact
affects significantly the behaviour of the blowup function and, consequently,
modifies the generic expression of the contact terms with respect to that 
of the pure gauge case, the latter being given by \lns
\eqn\confin{
{\CT}_{k,l} = - {1 \over 2\pi i} {\partial u_k \over \partial a^i}
{\partial u_l \over \partial a^j} ~ {\partial~ \over \partial\tau_{ij}}
\log \Theta_{[E]}(0|\tau) ~,}
where $[E]$ is a (non-singular) even half-integer characteristic, $[E] = 
[\vec\Delta,\vec 0]$. We further show how the methods of \bu\ can be 
extended to actually compute the polynomials 
${\cal B}^{(n)}_{i_1 \cdots i_n}(u_k,m_f,\Lambda)$ up to arbitrary order.

\subsec{Vanishing properties of $\Theta$--functions}

In contrast to what happens in the pure gauge case, the blowup function in 
the twisted theory with matter hypermultiplets has a fixed --and very
peculiar-- characteristic \betach\alphach. It is 
well known from the theory of Riemann surfaces (see for example \kra\belo) 
that characteristics can be associated to the branch points of the curve
through the elements of the Jacobian constructed out of the Abel map. More
concretely, consider a hyperelliptic curve $\Sigma_g$ of genus $g$ and a 
basis of homology cycles $(A^i,B_j) \in H_1(\Sigma_g,\IZ)$ with its 
respective normalized holomorphic differentials $d\omega_k$. Let 
$e_\alpha,~\alpha=1,\cdots,2g+2$ be the branch points of the surface, 
and take $e_1$ as a reference point. We can now define $2g+2$ vectors in 
the Jacobian $\vec{\cal U}_\alpha$ as the image of the divisors 
${\cal D}_\alpha = e_\alpha - e_1$ under the Abel map
\eqn\abelmap{
\vec{\cal U}_\alpha = {1 \over 2\pi i} \int_{e_1}^{e_\alpha} d\vec\omega =
\vec\epsilon_\alpha + \tau \vec\delta_\alpha ~,}
and the corresponding characteristic is $[{\cal U}_\alpha] = 1/2 
\left[\vec\epsilon_\alpha,\vec\delta_\alpha \right]$.

\bigskip
\centerline{\vbox{\hsize=3in\tenpoint
\centerline{\psfig{figure=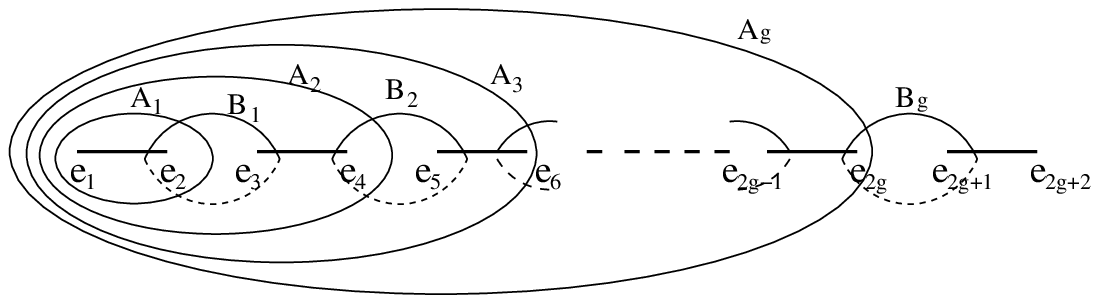}}
\vglue.2in
Fig. 1. Symplectic basis of homology cycles.}}
\vskip 3mm

There is a one to one identification among half-integer characteristics and
partitions of the branch points into two groups $I = I_m \cup (I/I_m)$,
where $I_m = \{e_{\alpha_1},e_{\alpha_2},\cdots,e_{\alpha_{g+1-2m}}\}$. It 
is given by
\eqn\assoc{
[\Upsilon] = \sum_{k=1}^{g+1-2m}[{\cal U}_{\alpha_k}] + [K] ~~~ 
({\rm mod} ~1) ~,}
where $[K]$ is the characteristic corresponding to the vector of Riemann 
constants $\vec K$ (which, in the case of a hyperelliptic curve, is a
half--period). For example, using the basis of cycles shown in Fig.1, the 
characteristic $[E]$ in \confin\ is the one corresponding to the
natural partition of branch points in the pure gauge Seiberg--Witten
solution. Finally, the characteristic of the vector of Riemann constants in
such basis,
\eqn\rc{
[K] = [\vec\Delta,\vec\Sigma] ~,}
turns out to be precisely the half-integer characteristic appearing in the 
blowup function of twisted theories with matter. For hyperelliptic curves
of genus $g>2$, this characteristic is {\it singular} having a zero of 
order $(g+1)/2$ or $g/2$ at the origin, provided $g$ is respectively odd 
or even. Then, the expansion \expansion\ of the blowup function in twisted 
theories with matter and gauge group $SU(2\ell)$ takes the form
\eqn\expmasa{
\tau_{\vec\Sigma}(t_i|u_k) = \sum_{n\ge \ell} ~\sum_{i_1, \cdots, i_n}
t_{i_1} \cdots t_{i_n} {\cal B}^{(n)}_{i_1 \cdots i_n} (u_k,m_f,\Lambda) ~.} 
Notice that, in contrast to the case with no flux, this blowup
function vanishes  at $t_i \to 0$.

\subsec{The hyperelliptic fundamental $\sigma$--function}

In order to further characterize the blowup function of twisted theories
with matter, let us introduce shortly some algebro--geometrical ingredients.
Given a hyperelliptic curve $y^2 = f(x)$ and a basis of Abelian 
differentials of the first kind $dv_k = x^{g-k}dx/y$, one can compute the 
period integrals
\eqn\periods{
A^i{_k} ={1\over 2\pi i } \oint_{A^i}dv_k ~, ~~~~~~~~~~~~~ 
B_{ik} ={1\over 2\pi i } \oint_{B_i}dv_k ~,}
in terms of which we can define the period matrix as 
\eqn\gaugeco{
\tau_{ij}=B_{ik} (A^{-1})^k_{\,\ j} ~.}
The low-energy ${\cal N}=2$ theory with $N_f$ matter hypermultiplets in the
fundamental representation is described by a prepotential ${\cal 
F}(a^i,m_f,\Lambda)$, where
\eqn\laa{
a^i(u_k,\Lambda) = {1 \over 2\pi i} \oint_{A^i} {x W'(x)\over \sqrt{W^2(x) 
- 4 \Lambda^{2N-N_f}}} ~dx ~,}
with $W = P/\sqrt{F}$, $P$ and $F$ being given in \chpol\ and \funf. The 
same expresion holds for the dual variables $a_{D,i} \equiv \p {\cal F}/\p 
a^i$, with $B_{i}$ instead of $A^i$. The effective gauge couplings are
given by \gaugeco, and
\eqn\ders{
A^i{_k} = {\partial a^i\over \p u_{k+1}} ~, \,\,\,\,\,\,\,\,\,\,\,\,\,\
B_{ik} = {\partial a_{D,i} \over \p u_{k+1}} ~.}

To construct a hyperelliptic $\sigma$--function, we also need a basis of 
Abelian differentials of the second kind $dr^k (x)$. It is provided by 
means of a Weierstrass polynomial $F(x_1,x_2)$, through the following 
identity \victoruno:
\eqn\definerre{
\sum_{k=1}^g dv_k(x_1) \, dr^k(x_2) = -{1 \over 2 y_1} {\partial \over
\partial x_2} \Bigl( {y_2 \over x_1-x_2}\Bigr) dx_1\, dx_2 + {F(x_1,x_2)
\over 4(x_1-x_2)^2} {dx_1\, dx_2 \over y_1 y_2} ~.}
Then, we define the following matrices of $\eta$--periods:
\eqn\etaperiods{
\eta^{ki} =-{1\over 2\pi i}\oint_{A^i}dr^k ~,~~~~~~~~~~~~~ \eta'^k_{\,\,\ i}
= -{1\over 2\pi i}\oint_{B_i}dr^k ~,}
that obey the Legendre relation
\eqn\legendre{
\eta = 2 \kappa A ~, ~~~~~~~~~~~~~~~ \eta' = 2 \kappa B - \half (A^{-1})^t ~,}
where $\kappa$ is a symmetric matrix that, of course, depends on $F$. These 
ingredients are enough to define the {\it hyperelliptic fundamental} 
$\sigma$--{\it function} by the formula \victoruno:
\eqn\sigm{
\sigma_f^F(\vec v) = C^{-1} ~\exp\{v_i \kappa^{il} v_l\} 
~\Theta_{[K]}((2\pi i)^{-1} v_l (A^{-1})^l_{\,\ i}|\tau) ~,}
where $C$ is constant (with respect to the $v_l$) and it is given by
\eqn\const{
C = i^{\ell} ~({\rm det} A)^{1/2}~ \Delta^{1/8} ~.}
It is now immediate to see that, after the identification $v_l \equiv 
it_{l+1}$, this is nothing but the blowup function of ${\cal N}=2$ twisted 
theories with gauge group $SU(2\ell)$ and $N_f < 4\ell$ matter 
hypermultiplets, {\it i.e.}, 
\eqn\finid{
\tau_{\vec\Sigma}(t_i|u_k) = i^{\ell} \sigma_f^F (v_l=it_{l+1}) ~.}
The overall factor $i^{\ell}$ in \const\finid\ is choosen just for
convenience.

The contact terms are given by a given matrix $\kappa$
(notice that their transformation properties under the action of the
modular group ${\rm Sp}(2g,\IZ)$, given in \victoruno\ and 
\egrmm, is indeed the same) for a given Weierstrass polynomial still
to be determined. In the pure gauge theory with magnetic fluxes turned off, 
the semiclassical vanishing of the contact terms allows for an analytical
determination of $F$. We shall study this issue in presence of matter in
the next section and see that there is no such a simplification and, in
turn, comparison with semiclassical results would become necessary. It is
somehow an expected result that the blowup function be a
$\sigma$--function, as long as the main property of the latter is its
invariance under the action of the modular group.

\subsec{Expansion of the blowup function}

The explicit expansion of the blowup function can be done following the 
method introduced in \bu, which is based in a series of developments 
carried out by Oskar Bolza one century ago \bolza. The starting point is
a partial differential equation for $\sigma_f^F$ with respect to a branch 
point $e_\alpha$ of $\Sigma_g$ \foot{As explained in \bu, there is a 
well-defined procedure to trade this sort of derivatives, which are of 
little practical use, for a differential equation involving 
$v_l$--derivatives and the coefficients of the curve.}, that can 
be written for any genus \bolza,
\eqn\bolzadifeq{
\eqalign{
{\partial {\sigma_f^F} \over \partial e_\alpha} + \sigma_f^F {\partial\log
C \over \partial e_\alpha} = & - \sum_{i,j=1}^g \biggl\{ p^F_{ij}(e_\alpha) 
v_i {\partial {\sigma_f^F} \over \partial v_j} + {1 \over 2} {\sigma_f^F} 
q^F_{ij}(e_\alpha) v_i v_j \cr & ~~~~~~~~~~~~ - {e_\alpha^{2g-i-j} \over 
f'(e_\alpha)} \Bigl( {\partial^2{\sigma_f^F} \over \partial v_i \partial 
v_j} - 2 \kappa_{ij} ~{\sigma_f^F} \Bigr) \biggr\} ~,}}
where the matrices $p^F_{ij}(e_\alpha)$ and $q^F_{ij}(e_\alpha)$ are given by
\eqn\matri{
\eqalign{
\sum_{i,j=1}^g p^F_{ij}(e_\alpha) x^{g-i} h_j (z) = & ~{1\over 2}{(x-z)^{g-1}
\over x-e_\alpha}-{1\over 2}{(e_\alpha-z)^{g-1} \over f'(e_\alpha)} 
{F(x,e_\alpha) \over (x-e_\alpha)^2} ~,\cr
\sum_{i,j=1}^g q^F_{ij}(e_\alpha) x^{g-i} z^{g-j} = & ~{1\over 8}\Bigl({1
\over x-e_\alpha} + {1 \over z-e_\alpha}\Bigr){F(x,z) \over (x-z)^2} + 
{1 \over 4}{1 \over (x-z)^2}{\partial F(x,z)\over \partial e_\alpha} \cr 
& ~~~- {1\over 8} {F(x,e_\alpha) F(z,e_\alpha) \over f'(e_\alpha)
(x-e_\alpha)^2 (z-e_\alpha)^2} ~,}}
the $'$ denoting derivatives w.r.t. $x$, and the function $h_j(z)$ being 
implicitely defined through the relation
\eqn\haches{
(x-z)^{g-1}=\sum_j x^{g-j}h_j(z) ~.}
If we plug in the Taylor expansion of $\sigma_f^F$ in \bolzadifeq,
\eqn\taylorexp{
\sigma_f^F(\vec v) = \sum_{n=\ell}^\infty ~\varsigma_n(\vec v) ~,}
where $\{\varsigma_n(\vec v)\}$ are homogeneous polynomials of degree $n$
in $v_l$ (notice that the sum runs over even or odd integers according 
to the parity of $\ell$), a set of recursive relations shows up immediately:
\eqn\recursive{\eqalign{
{\partial \varsigma_{n-2} \over \partial e_\alpha} + \varsigma_{n-2}
{\partial \log C \over \partial e_\alpha} = & - \sum_{i,j=1}^g 
\biggl\{ p^F_{ij}(e_\alpha) v_i {\partial \varsigma_{n-2} \over \partial 
v_j} + {1 \over 2} q^F_{ij}(e_\alpha) v_i v_j ~\varsigma_{n-4} \cr &
~~~~~~~~~~~~~~~~~ - {e_\alpha^{2g-i-j} \over f'(e_\alpha)}\Bigl( {\partial^2 
\varsigma_{n} \over \partial v_i \partial v_j} - 2\kappa_{ij}
~\varsigma_{n-2} \Bigr) \biggr\} ~.}}
We already know that the leading term in $\sigma_f^F$ at the origin is of 
order $\ell$. Then, setting $n = \ell$ in \recursive, we obtain
\eqn\priter{
\sum_{i,j=1}^g e_\alpha^{2g-i-j} {\partial^2\varsigma_{\ell} \over
\partial v_i \partial v_j} = 0 ~,}
for any branch point $e_\alpha$. The solution to this equation is provided 
by the determinant of the Hankel matrix
\eqn\hankel{
\varsigma_{\ell} = \det H = \det\pmatrix{
v_1      & v_2        & \ldots & v_{\ell}    \cr
v_2      & v_3        & \ldots & v_{\ell+1}  \cr
\vdots   & \vdots     & \ldots & \vdots      \cr
v_{\ell} & v_{\ell+1} & \ldots & v_{2\ell-1}} ~.}
The overall multiplicative factor has been fixed by comparison with the 
result obtained in \victoruno\ following a different approach. Then, 
the leading term of the blowup function is given by $i^{\ell}$ times 
the determinant of the Hankel matrix with $v_l \to it_{l+1}$. This is the
answer for ${\cal B}^{(\ell)}_{i_1 \cdots i_{\ell}} (u_k,m_f,\Lambda)$. 
Notice that the constant $C$ is nothing but
\eqn\constag{
C = {\partial^{\ell} ~\Theta_{[K]}((2\pi i)^{-1} v_l (A^{-1})^l_{\,\
i}|\tau) \over \partial v_1 \partial v_3 \cdots \partial v_{2\ell-1}}
\bigg|_{\vec v = \vec 0} ~.}
Being given by the derivative of a $\Theta$--function, the derivative of
$\log C$ with respect to $e_\alpha$ can be treated again by means of the
formalism developed by Bolza \bolza\ resulting in
\eqn\otrab{
{\partial \log C \over {\partial e_{\alpha}}} = \sum_{i,j=1}^g
{e_\alpha^{2g-i-j} \over f'(e_\alpha)} \biggl\{ {\partial^{\ell+2} 
\sigma_f^F \over \partial v_i \partial v_j \partial v_1 \cdots \partial 
v_{2\ell-1}} \bigg|_{\vec v = \vec 0} - 2 \kappa_{ij} \biggr\} -
\sum_{i=1}^g p_{ii}^F(e_{\alpha}) ~.}
Then, the recursive relations \recursive\ are
\eqn\recurfin{\eqalign{
{\partial \varsigma_{n-2} \over \partial e_\alpha} = & \sum_{i=1}^g 
p_{ii}^F(e_{\alpha}) ~\varsigma_{n-2} - \sum_{i,j=1}^g \biggl\{ 
p^F_{ij}(e_\alpha) v_i {\partial \varsigma_{n-2} \over \partial v_j} + 
{1 \over 2} q^F_{ij}(e_\alpha) v_i v_j ~\varsigma_{n-4} \cr &
~~~~~~~ - {e_\alpha^{2g-i-j} \over f'(e_\alpha)}\Bigl( {\partial^2 
\varsigma_{n} \over \partial v_i \partial v_j} - {\partial^{\ell+2} 
\sigma_f^F \over \partial v_i \partial v_j \partial v_1 \cdots \partial 
v_{2\ell-1}} \bigg|_{\vec v = \vec 0} ~\varsigma_{n-2} \Bigr) \biggr\} ~.}}
Notice that the coefficient of the last term in the second line only
depends on $\varsigma_{\ell}$ and $\varsigma_{\ell+2}$ so that 
recursivity is safe.

The result obtained above is nontrivial in the sense that it is not 
dictated {\it a priori} by symmetry arguments or the semiclassical behaviour. 
Provided the Weierstrass polynomial is obtained for a given gauge group and 
matter content, the recursive rule \recursive\ would allow for a
computation of the blowup function up to arbitrary order in time
variables. Finally, once the blowup function is recognized to be a 
hyperelliptic $\sigma$--function, it is immediate to show that it 
is the $\tau$--function of a finite gap solution of the KdV hierarchy. In
fact, the proof is exactly the same given in \bu\ for the pure gauge theory.

\newsec{Contact terms}

As originally noticed in \lns, an explicit expression for the contact terms 
can be derived from the blowup function by requiring invariance under
$Sp(2g,\IZ)$ duality transformations, and taking into account that they
must vanish semiclassically \mw. We shall perform a similar analysis in
what follows. It is instructive to consider first the case of $SU(2)$, 
where everything can be written in terms of elementary elliptic functions. 
In this case, the blowup function \blowupf\ is given by 
\eqn\sutwo{
\tau_{1 \over 2}(t_i|u_k) = h^{-1/2} ~\Delta^{-1/8} ~{\rm e}^{-t^2{\cal 
T}_{2,2}} ~{\vartheta_1 ((2\pi h)^{-1}t|\tau)} ~, } 
where $h=da/du$. Using Thomae's formula, and choosing the 
appropriate normalization, one finds that $\Delta^{1/8} =(1/2) h^{-3/2} 
\vartheta_2(0|\tau) \vartheta_3(0|\tau) \vartheta_4(0|\tau)$. Using the 
identity $\vartheta'_1 (0|\tau) = - \pi \vartheta_2(0|\tau) 
\vartheta_3(0|\tau) \vartheta_4(0|\tau)$, the expansion of \sutwo\ in $t$
turns out to be
\eqn\exp{
\tau_{1 \over 2}(t_i|u_k) = - t \Bigl\{ 1 - t^2 \Bigl( {\cal 
T}_{2,2} - { 1\over 24 \pi^2 h^2} {\vartheta'''_1 (0|\tau) \over 
\vartheta'_1 (0|\tau)} \Bigr) + {\cal O}(t^4)\Bigr\} ~.}
It follows from \expmasa\ that the contact term is
\eqn\consutwo{
{\cal T}_{2,2} = {1\over 24 \pi^2 h^2} {\vartheta'''_1 (0|\tau) \over
\vartheta'_1 (0|\tau)} +  ~{\cal B}^{(3)}(u,m_f,\Lambda) ~.}
As discussed above, ${\cal B}^{(3)}(u,m_f,\Lambda)$ is such that 
the contact terms vanish semiclassicaly. Taking into account the identity
\eqn\id{
{\vartheta'''_1 (0|\tau) \over \vartheta'_1 (0|\tau)} = - \pi^2 E_2(\tau) ~,}
where $E_2(\tau)$ is the normalized Eisenstein series, we see that 
\consutwo\ is indeed consistent with the explicit expression for the 
contact terms derived in \mw\lns\mmone: 
\eqn\ctdos{
{\cal T}_{2,2} = - {1 \over 24 h^2} E_2(\tau) + {1 \over 3} \biggl( u +
\delta_{N_f,3} {\Lambda^2 \over 64} \biggr) ~,}
and from this equation we can also read the 
explicit value of ${\cal B}^{(3)}(u,m_f,\Lambda)$. Now, as mentioned 
earlier, the contact term ${\cal T}_{2,2}$ in such case is given by
\confin\ with $k=l=2$. Thus, the following identity emerges:
\eqn\thetaide{
-{1 \over 2\pi i} \Bigl( {du \over da} \Bigr)^2 \partial_{\tau} \log
\vartheta_4 (0|\tau) = {1\over 24 \pi^2} \Bigl( {du \over da} \Bigr)^2
{\vartheta'''_1 (0|\tau) \over \vartheta'_1 (0|\tau)} + {u \over 3} ~.}
This result can be shown analytically by means of the theory of elliptic
functions \lns. As we will see in what follows, different expressions for
the contact terms lead to generalizations of this sort of identities to 
higher genus curves.
  
Let us first introduce a short notation for the contraction of derivatives
of the singular $\Theta$--function and the inverse of the $A^i$--periods 
of the Seiberg--Witten differential
\eqn\defvth{
\vartheta^{(n)}_{i_1\cdots i_n} \equiv {1 \over (2\pi)^n n!} ~{\partial 
u_{i_1}
\over \partial a^{j_1}} \cdots
{\partial u_{i_n} \over \partial a^{j_n}} ~ 
{\partial^n \Theta_{[K]}(0|\tau) \over
\partial z_{j_1}  \cdots \partial z_{j_n}} ~.}
Now, taking into account our earlier results, the expansion of 
$\Theta_{[K]}(\vec z\mid\tau)$ reads
\eqn\taylor{
\Theta_{[K]}(\vec z\mid\tau) = t_{i_1} \cdots t_{i_{\ell}} 
~\vartheta^{(\ell)}_{i_1 \cdots i_{\ell}} + t_{i_1} \cdots t_{i_{\ell+2}} 
~\vartheta^{(\ell+2)}_{i_1 \cdots i_{\ell+2}} + \cdots ~,} 
where repeated indices are summed. If we now expand the blowup function 
\blowupf\ 
and we compare the result with the structure of \expmasa, we find the 
following equations for the contact terms:
\eqn\conterm{
{i^{\ell} \over C}\vartheta^{(\ell)}_{(i_1 \cdots i_{\ell}}{\cal T}_{mn)} = 
{i^{\ell} \over C}\vartheta^{(\ell+2)}_{i_1 \cdots i_{\ell}mn}-
{\cal B}^{(\ell+2)}_{i_1 \cdots i_{\ell}mn} (u_k,m_f,\Lambda) ~,}
where the indices in the left hand side of the equation are symmetrized. 
This is, in principle, the generalization of \consutwo\ to the higher 
rank case. One can also obtain a much more explicit equation for 
${\cal T}_{2,2}$ by putting $t_{i>2} \equiv 0$, and considering only terms 
in the blowup function depending on $t_2$. Since 
$\vartheta^{(\ell)}_{2\cdots 2}$ {\it vanishes} due to
the fact that the determinant of the Hankel matrix has no $v_1^{\ell}$
term, we need to attain higher orders in the expansion of the blowup 
function. The first nonvanishing term in the expansion is the 
$v_1^{\ell^2}$ term. This is due to the fact that 
$\varphi^F$ is a homogeneous function of degree $-\ell^2$ provided we 
assign a weight equal to $-i$ for $v_i$. The 
expression for the contact term ${\cal T}_{2,2}$ arising from this 
analysis is
\eqn\tdd{
{\cal T}_{2,2} =
{\vartheta^{(\ell^2+2)}_{2\cdots 2}\over 
\vartheta^{(\ell^2)}_{2\cdots 2}}
 - {{\cal B}^{(\ell^2+2)}_{2 \cdots 2}(u_k,m_f,\Lambda)\over {\cal 
B}^{(\ell^2)}_{2 \cdots 2}(u_k,m_f,\Lambda)} ~.}
For the general case of $t_{i>2} \neq 0$, \conterm\ gives different
expressions for the contact terms involving derivatives of higher Casimir
operators. 

To illustrate the above results, consider the 
case of pure gauge $SU(4)$ theory. The 
Weierstrass function that appears in the 
$\sigma$--function has the form \bu\
\eqn\wei{
F = Q(x_1)R(x_2) + Q(x_2)R(x_1) ~,}
where the polynomials $Q$ and $R$ are given by
\eqn\polqyp{
Q(x) = P(x) + 2 \Lambda^4 ~, ~~~~~~~~~~~~R(x) = P(x) - 2\Lambda^4 ~.}
Using the differential equations of section 3, one obtains the 
expansion
\eqn\suc{\eqalign{
\sigma^F_f & = t_3^2 - t_2 t_4 - {1 \over 12} t_2^4 + {u_2 \over 6} t_4 t_2^3 -
{u_3 \over 3} t_2 t_3^3 + {u_3 \over 2} t_2^2 t_3 t_4 - {u_2 u_3 \over 6}
t_3^3 t_4 - {u_3^2 \over 8} t_3^2 t_4^2 + {u_4 \over 2} t_2^2 t_4^2 \cr
& - {u_3 u_4 \over 6} t_3 t_4^3 - {u_2^2 + 4 u_4 \over 12} t_3^4 + {u_3^2 -
4 u_2 u_4 \over 24} t_2 t_4^3 + {4 \Lambda^8 - u_4^2 \over 12} t_4^4 + {u_2
\over 180} t_2^6 ~.}} 
We then find ${\cal
B}^{(6)}_{2 \cdots 2}/{\cal
B}^{(4)}_{2 \cdots 2} = - {u_2 \over 15}$. One can then check 
\tdd\ by using the semiclassical expansion of the effective gauge coupling 
up to order $\Lambda^8$. This result for the quotient of the 
${\cal B}$'s not also holds in the $N_f=1$ and $N_f=2$ theories. 
As a further check, one finds that for $SU(6)$ ${\cal 
B}^{(11)}_{2 \cdots 2}/{\cal 
B}^{(9)}_{2 \cdots 2} = - {3 \over 70} u_2$.

The expressions obtained above for the contact terms are valid for any
number of massive hypermultiplets. In particular, they apply when $N_f$ is an
even number and the bare masses are degenerated in pairs, $m_f = 
m_{f+N_f/2}$. In such cases, the results for the contact terms must 
coincide with those carried out earlier in \egrmm. For example, comparison 
of ${\cal T}_{2,2}$ implies the following identity between hyperelliptic
$\Theta$--functions,
\eqn\relt{
-{1 \over 2\pi i} {\partial u_2 \over \partial a^i} {\partial u_2 \over 
\partial a^j} {\partial~ \over \partial \tau_{ij}} \log \Theta_{[E]}(0|\tau) =
 ~{~\vartheta^{(\ell^2+2)}_{2\cdots 2}\over 
\vartheta^{(\ell^2)}_{2\cdots 2}}
 - {{\cal B}^{(\ell^2+2)}_{2 \cdots 2}(u_k,m_f,\Lambda)\over 
{\cal B}^{(\ell^2)}_{2 \cdots 2}(u_k,m_f,\Lambda)}.}
Similar expressions can be obtained in principle from \conterm.

\medskip

The new expressions for the contact terms found in the present paper should
be useful to compute the instanton
corrections to the effective prepotential of ${\cal N}=2$ supersymmetric
theories with arbitrary matter content. Since the second derivative 
of the prepotential $\partial^2 {\cal F} / \partial \Lambda^2$ 
is (up to a numerical constant) equal to ${\cal T}_{2,2}$, and this 
is given by \tdd, one 
can follow the arguments in \emm\egrm\edemas\ to obtain the semiclassical 
expansion of ${\cal F}$ through a recursive procedure. 


\bigskip
\noindent
{\bf Acknowledgements:} 
The work of J.D.E. has been supported by the Argentinian National Research 
Council (CONICET) and by a Fundaci\'on Antorchas grant under project number 
13671/1-55. The work of M.M. has been supported by DOE grant 
DE-FG02-96ER40959.  

\listrefs

\bye

\listrefs

\end